# Integrated photonic building blocks for next-generation astronomical instrumentation I: the multimode waveguide

**Nemanja Jovanovic,**[1,2,3*] **Izabela Spaleniak,**[1,2] **Simon Gross,**[1,4] **Michael Ireland,**[1,2,3] **Jon S. Lawrence,**[1,2,3] **Christopher Miese,**[1,4] **Alexander Fuerbach,**[1,4] **and Michael J. Withford**[1,2,4]

[1]*MQ Photonics Research Centre, Dept. of Physics and Astronomy, Macquarie University, NSW 2109, Australia*
[2]*Macquarie University Research Centre in Astronomy, Astrophysics & Astrophotonics, Dept. Physics and Astronomy, Macquarie University, NSW 2109, Australia*
[3]*Australian Astronomical Observatory (AAO), PO Box 296, Epping NSW 1710, Australia*
[4]*Centre for Ultrahigh Bandwidth Devices for Optical Systems (CUDOS), Australia*
[*]*jovanovic.nem@gmail.com*

**Abstract:** We report on the fabrication and characterization of composite multimode waveguide structures that consist of a stack of single-mode waveguides fabricated by ultrafast laser inscription. We explore 2 types of composite structures; those that consist of overlapping single-mode waveguides which offer the maximum effective index contrast and non-overlapped structures which support multiple modes via strong evanescent coupling. We demonstrate that both types of waveguides have negligible propagation losses (to within experimental uncertainty) for light injected with focal ratios >8, which corresponds to the cutoff of the waveguides. We also show that right below cutoff, there is a narrow region where the injected focal ratio is preserved (to within experimental uncertainty) at the output. Finally, we outline the major application of these highly efficient waveguides; in a device that is used to reformat the light in the focal plane of a telescope to a slit, in order to feed a diffraction-limited spectrograph.

## 1. Introduction

Over the past three decades there has been an increase in the incorporation of photonic technologies into world-class astronomical observatories which prompted the branding of a new field: "astrophotonics" in 2009 [1]. This has been driven by the increasing observational demands placed on observatories by the astronomy community. An ideal example is the push for large-scale multi-object spectroscopic surveys in the early 2000's. Such surveys required the collection of spectra from hundred's of thousands of galaxies which was not feasible with the technology at the time in any reasonable timeframe. Instrument scientists opted to utilize multimode fibers to collect the light from hundreds of galaxies simultaneously from the focal plane and reformat the fibers into a slit at the input of a large spectrograph that could now be located in an environmentally stable room away from the telescope, a concept first demonstrated 20 years earlier [2]. By robotically controlling the position of the fibers, astronomers were able to collect thousands of spectra a night, increasing their observational efficiency by several orders of magnitude, paving the way for an era of large-scale multi-object surveys [3,4].

Photonic technologies, including both integrated and fiber-based platforms offer numerous advantages. For instance in the example above the small footprint of the multimode fibers, coupled with their flexibility and robustness made them the ideal choice for that application. The main benefit of the integrated platform is that it offers the possibility to concatenate many photonic components with various functionalities onto a miniature monolithic wafer/chip consisting of only a handful of materials [5,6]. This leads to a mechanically robust instrument that offers superior resistance to environmental influences such as temperature. From an optical standpoint the greatest benefits of photonic functionalities are obtained when operating in the single-mode regime. The first of these is spatial filtering of the stellar signal, which increases the signal-to-noise ratio of fringe contrasts in stellar interferometry [7]. The second and more relevant advantage to this work is that the footprint of a spectrograph for a given resolving power is much smaller when diffraction-limited light is injected as opposed to atmospherically perturbed light (which cannot be focused to the diffraction-limit and is known as seeing-limited) [8]. A reduction in size has a direct bearing on the robustness to flexure and overall stability of the spectrograph.

To this day many astronomical observatories do not exploit any type of wavefront correction such as adaptive optics (AO), prior to injecting into their spectrographs (these are known as seeing-limited spectrographs). Even when wavefront correction is available with large sky coverage, Strehl ratios (ratio of the observed peak intensity at the focal plane from a point source as compared to the theoretical maximum peak intensity of a diffraction limited imaging system) are typically low for wavelengths shorter than 2 μm. For such spectrographs it is not possible to efficiently couple the light from the atmospherically induced speckle pattern in the focal plane of the telescope into a single-mode or even a few-moded fiber for two reasons: the étendue of the beam (*spot size × speed of the beam*) cannot be preserved when coupling into such fibers and there is a poor spatial overlap between the high spatial frequency speckle pattern and the lowest few modes of the fibers [9,10]. Instead a multimode (MM) fiber that can preserve the étendue of the beam and supports sufficient modes to sample the speckle pattern appropriately is required for efficient coupling from the focal plane. A subsequent transition to convert to a single-mode (SM) format is then needed. Such a transition was proposed [11] and demonstrated at 1.55 μm (astronomical H-band) and is known as the photonic lantern [12,13]. Photonic lanterns consist of an array of SM fibers that are adiabatically tapered into a single MM fiber. If the number of non-degenerate modes in the MM section matches the number of SM fibers, then the degrees of freedom of the system are maintained and it is possible to efficiently convert the seeing-limited light into a diffraction-limited format albeit with the signal now split amongst numerous optical fibers. However, if the fibers are all aligned along a slit and injected into a spectrograph and provided that the signal is above the noise, then it is possible to add the signal from each spectrum during post processing [14].

The majority of the spectral lines in astronomically relevant targets at low redshift (close to us) are in the visible spectral region and moving to these short wavelengths means a dramatic increase in the number of modes supported by the MM section (which scales as $1/\lambda^2$). Hence, a corresponding increase in the number of SM fibers is required. Once thousands of SM fibers are required it becomes logistically difficult to handle and thus it would be ideal to embed them into a monolithic, integrated photonic chip. A prototype integrated photonic lantern was realized by Thomson *et al.* in 2011 [15]. It was fabricated by ultrafast laser inscription (ULI) of the structure into a glass substrate; one of the only techniques which allow true 3D architectures to be realized; a requirement for this kind of device. Although the performance of the prototype was encouraging, a more detailed study which takes into consideration the spot size and speed of the beam being injected into such a structure at the telescope facility is required, in order to develop a "recipe" for creating optimized ultrafast laser-inscribed transitions.

Even before the MM to SM transition and its design characteristics are to be considered, the first element of the device, namely the MM waveguide must thoroughly be investigated. The fabrication of large waveguides by ULI capable of supporting multiple modes was first demonstrated in 2003 by Nolte *et al.* [16]. In 2010, we undertook (to the best of our knowledge) the first attempt to optimize and characterize ultrafast laser inscribed MM waveguides. Those guides consisted of several large overlapping tracks of index change [17].

Cross-sectional images of two waveguides fabricated in that study are shown in Fig. 1(a). Recently we discovered that some of these waveguides have discontinuities within the sample as seen in Fig. 1(b), an effect brought on by the high average powers (>2W, 5.1 MHz repetition rate, 400 nJ per pulse) required to drive the cumulative heating process in order to create such large guides. This in turn boiled the index matching oil between the writing objective and the sample, causing the waveguide to terminate abruptly. Such structural irregularities impact the throughput of the waveguides and thus preclude them from use in astronomical applications in their current state. It should be noted that this, at least to some extent, could be overcome by using a non-index matched writing system. However, in this body of work we shift our focus to an alternative approach whereby we construct composite MM waveguides from a lattice of 10's of SM waveguides each written with low average powers instead. In section 2 we outline the fabrication and characterization procedures while section 3 highlights the performance of both overlapping and non-overlapping composite structures. In section 4 we outline the primary application of the composite guides, which is in slit reformating devices for a diffraction-limited spectrographs and describe in detail the procedure for designing devices based on these guides.

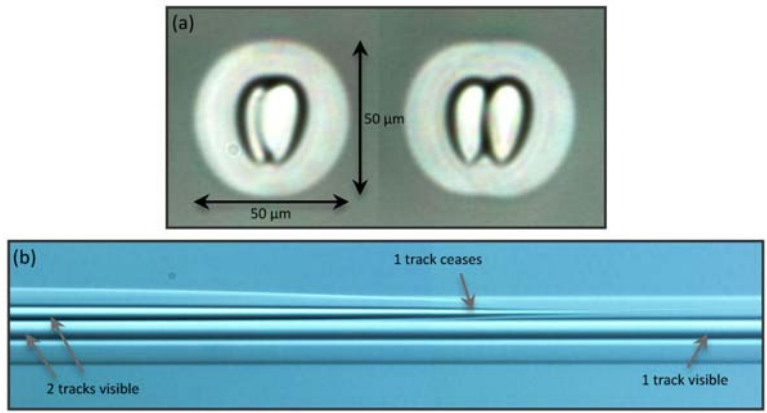

Fig. 1. (a) Cross-sectional micrograph of two MM waveguides fabricated from 2 laser scans (tracks) with spacing's of 5 μm (Left) and 10 μm (right). (b) Micrograph image taken from the top of the MM waveguide with the 10 μm spacing. It can clearly be seen that the second track ceases within the chip, a byproduct of the index matching oil boiling during the writing process.

## 2. Waveguide fabrication and characterization

The MM waveguides were stitched together from a lattice of SM waveguides using the multiscan method [18]. They were inscribed using an ultrafast titanium sapphire oscillator (Femtolasers GmbH, FEMTOSOURCE XL 500, 800 nm centre wavelength, <50 fs pulse duration) with 5.1 MHz repetition rate. The laser was focused into a 30 mm long boro-aluminosilicate (Corning Eagle2000) glass sample using a 100× oil immersion objective lens (Zeiss N-Achroplan, numerical aperture ($NA$)=1.25, working distance = 450 μm). Pulse energies of 35 nJ were used in conjunction with translation velocities of 250-2000 mm/minute in order to create waveguides which were single-mode at 1550 nm. A set of Aerotech, air-bearing translation stages were used to smoothly translate the sample during writing, and in a piecewise fashion construct a lattice of single-mode guides which constituted the multimode guide. The chip was ground and polished to reveal the waveguide ends. As a result of the high repetition rates used all waveguides presented herein were written in the cumulative heating regime [19], on time-scales of a minute.

    The MM waveguides were characterized using the setup depicted in Fig. 2. A light emitting diode (LED) with a centre wavelength of 1550 nm and a full-width at half maximum

(FWHM) bandwidth of 115 nm was used as the probe source (Thorlabs-LED1550E). It was driven by a current source set to 60 mA. The light was coupled into a 50 μm core diameter, 0.22 *NA*, MM fiber which was coiled twice around a 30 mm diameter drum in order to uniformly illuminate the modes of the fiber. The output of the fiber was collimated by a 30 mm focal length achromat (Thorlabs-AC254-030-C-ML) before being injected into the MM waveguides by means of a 25 mm focal length achromat (Thorlabs-AC127-025-C-ML). In order to control the focal ratio (*F/#*)/*NA* of the injected beam a calibrated iris (Thorlabs-SM1D12C) was placed in the collimated beam. The MM fiber, the iris and the collimating and injection optic were all mounted on a 6-axis (x, y, z, $\theta_x$, $\theta_y$, $\theta_z$), flexure translation stage (Thorlabs-MAX601D/M), which allowed for fine adjustments of the coupling into the waveguides.

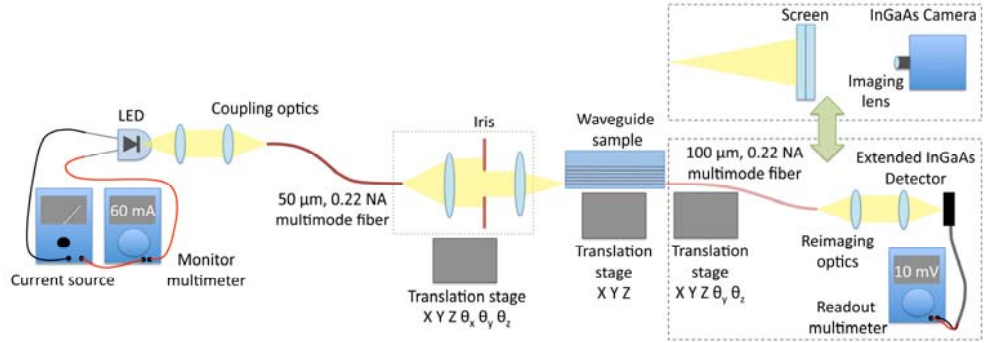

Fig. 2. Schematic diagram of the characterization setup for the MM waveguides.

In order to make throughput measurements a 100 μm core diameter, 0.22 *NA*, MM fiber was used to collect the light at the output of each waveguide. It was butt coupled to the photonic chip, immersed in index matching oil in order to eliminate Fresnel reflections and aligned with each waveguide by means of a 5-axis (x, y, z, $\theta_y$, $\theta_z$), flexure translation stage (Thorlabs-MAX313D/M+APY002/M). The output of the fiber was reimaged onto an extended InGaAs detector (Thorlabs-PDA10DT-EC), which was monitored by a multimeter. The throughputs of the waveguides were normalized to the amount of light that was collected by the 100 μm collection fiber when it was placed directly at the focus of the injection lens, with the photonic chip removed. By using this method a consistent number of air/glass interfaces, which cause Fresnel reflections, was maintained between the waveguide and calibrator measurements with the only difference being the presence of the waveguide itself. The normalized throughput was then rescaled to remove the effect of absorption by the Eagle2000 substrate, so that the true performance of the MM waveguides could be assessed. It was determined that the absorption coefficient of Eagle2000 glass at 1550 nm was $\alpha = 0.0075 \pm 0.0005$ mm$^{-1}$ and hence the post-polished chip which was $29.0 \pm 0.1$ mm in length was limited to a maximum transmission of $80.5 \pm 1.1\%$. It should be noted that this is the first time that the background absorption of Eagle2000 has been reported at 1550 nm. Indeed this background absorption entirely accounts for the lowest reported propagation losses of ~0.30 dB/cm at 1550 nm (to within experimental uncertainty), which did not consider this loss mechanism at the time [20]. Similarly from our own experiments in Eagle2000 at 1550 nm, we have determined that the losses in straight waveguides can be entirely attributed to the coupling and absorption losses in all cases. Propagation losses (losses due to the waveguide structure itself) in single-mode waveguides at 1550 nm in Eagle2000 are therefore negligible. This is an important clarification for the future of the field of ULI. As a final step, the throughputs were rescaled a second time to take into account coupling losses to first order. Although the FWHM size of the launch field was $46 \pm 2$ μm, it did not have the ideal flat-topped cross-sectional profile as can be seen in Fig. 3 below (red curve), and hence there was significant power outside the physical extent of the waveguides. This power did not couple into the waveguides and hence should not be included in our measurements of throughput. The launch field deviated even further from a flat-top at high *F/#s* (blue curve in Fig. 3), which was due to the fact that the iris was in a near-closed position and hence the spatial frequencies of the image of the injection MM fiber facet were filtered. All launch fields as a function of the *F/#* of the beam being injected were convolved with the flat-top profile which was circular in shape and slightly larger than the physical extent of the waveguides (grey dashed line in Fig. 3). This process is referred to as windowing from this point on while the flat-top profile is referred to as the window. The fraction of power in the launch field within the window was then used as the scaling factor to rescale the throughputs. Note that this circular shaped windowing function did not ideally match the hexagonal geometries presented below but is a good first approximation. All throughputs reported hereafter have been rescaled to compensate for both of the above-mentioned effects. Hence we are affectively reporting the waveguide propagation losses in isolation in this publication.

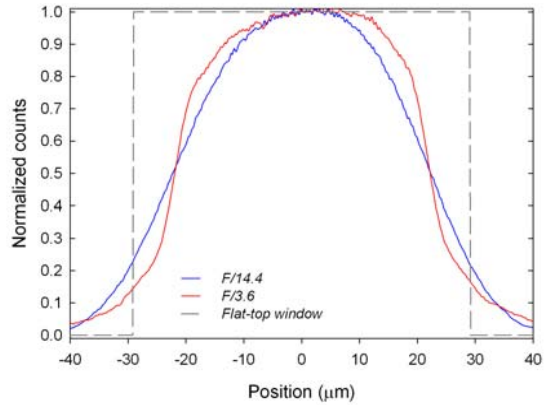

Fig. 3. Line profiles through the energy distributions of the launch field at the point of focus (input plane of the chip) for a *F/3.6* (red line) and *F/14.4* (blue line) beam. The flat-top window used to represent the approximate size of the waveguides (grey dashed line).

Measurements of the *F/#/NA* of the launch setup and the waveguides were carried out by substituting the fiber collecting assembly that is depicted in the bottom right of Fig. 2 with the imaging setup shown in the top right of Fig. 2. In this setup the light from the launch assembly or the waveguides was allowed to diffract over a distance $z \sim 20$ cm onto a screen formed by a sheet of diffuse tracing paper sandwiched between two 4-inch glass plates. This distance ensured that the spot on the screen was the far-field intensity distribution. An InGaAs, CMOS camera (Xenics-Xeva–1.7-640) was used to image the spot on the screen. Dark frames were initially subtracted from the data and then the 90% encircled energy spot size (diameter, *D*) was calculated for each image. The spot size was calibrated by means of a linear scale on a transparency slide that was also placed between the two glass plates. The *NA* was calculated from the following equation:

$$NA = \sin\left(\tan^{-1}\left(\frac{D}{2z}\right)\right). \quad (1)$$

The *F/#* was determined by then using $F/\# = 1/(2 \times NA)$.

## 3. Performance of composite multimode waveguides

In this section we will summarize the results for a range of composite MM waveguide structures. All structures were designed to be approximately 50 μm in diameter in order to match to conventional MM fibers and simplify the characterization process. The first type of waveguide to be discussed consisted of 37 SM guides (referred to as "tracks") placed sufficiently close together (8 μm pitch) in a hexagonal lattice such that they overlap with one another as seen in Fig. 4; named the "Hex 37" guide. This type of waveguide has the highest possible effective index as the entire cross-sectional area that the guides occupy has been modified. Subsequently we will explore more sparsely populated lattices that consist of 19 waveguides each, which do not overlap; named the "Hex 19" and "Circ 19" guides also shown in Fig. 4. These waveguides support multiple modes because the SM tracks, which they comprise of, are placed close enough together that they strongly couple. The primary advantage of such structures is the fact that the laser does not need to write over any region more than once, resulting in a higher degree of homogeneity amongst index modifications.

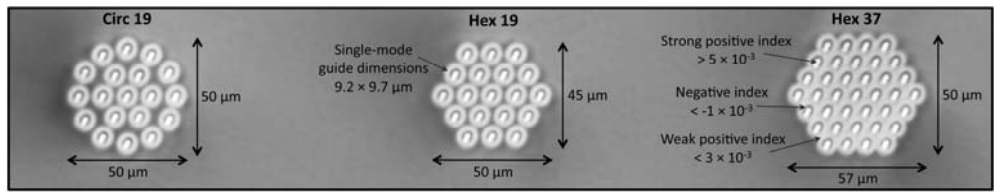

Fig. 4. Cross-sectional micrograph of the three types of composite MM waveguides fabricated and characterized. The SM waveguide tracks were written with 35 nJ and a translation speed of 1750 mm/minute. The dimensions of the waveguides written under these conditions are marked in the figure and the three main regions of index change, typically seen under cumulative heating in Eagle2000 glass, are highlighted for the reader.

*3.1 Overlapped multimode waveguide structures - throughput and cutoff focal ratio*

The normalized throughput for the set of Hex 37 waveguides as a function of the injected $F/\#$ is shown in Fig. 5(a). Figure 5(b) summarizes the normalized throughput as a function of the translation speed used for inscription. It can be seen that the throughput increases as a function of increasing $F/\#$ and flattens off after ~$F/8$ at ~100% (to within experimental uncertainty). The error bars are based on the uncertainty in the background absorption, the length of the chip, the accuracy of the size of the iris which controls the $F/\#$ as well as any drift in the calibration curves taken for the launch setup without the waveguides in the way, before and after the experiment. The increasing error bars as we move to higher $F/\#$s (in both Fig. 5(a) and (b)) are due to the smaller apertures of the iris which heavily attenuate the signal. As a result even the slightest changes in the calibration curve with time become a large fraction of the total signal at high $F/\#$s. For $F/\#>8$ the waveguides do not exhibit any propagation losses to within the ~7% experimental uncertainty. Interestingly, for $F/\#<8$ the throughputs increase by ~10% as the translation speed is decreased (see Fig. 5(b)). This arises because a reduction in translation speed leads to an increase in deposited energy per unit volume, which results in both a slightly larger guide but more critically, a stronger index modification ($\Delta n$). A stronger index modification is capable of guiding rays at higher angles of inclination to the optic axis (i.e. higher *NA*, low *F/#*).

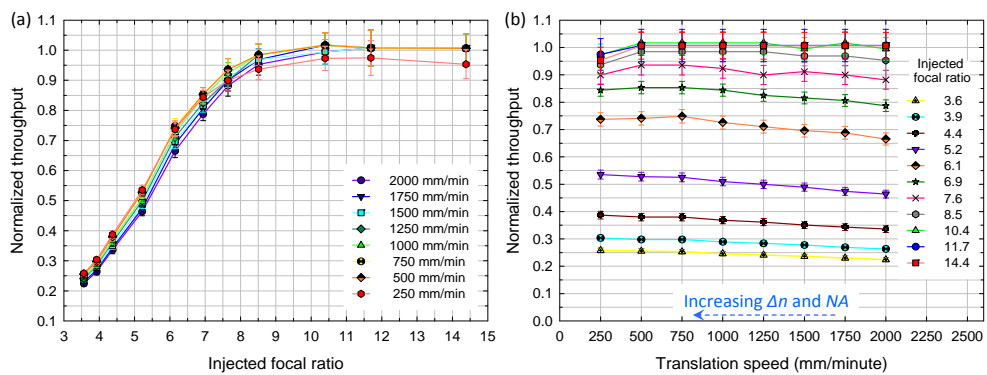

Fig. 5. (a) Normalized throughput for the Hex 37 waveguide structures written with different translation speeds as a function of the injected focal ratio. (b) Normalized throughput for the Hex 37 waveguide structures for various injected focal ratio's as a function of the translation speed used for inscription.

The far-field distributions for three of the Hex 37 waveguides were imaged and the data is reported in Fig. 6. Figure 6(a) and (b) show typical far-field energy distributions for the launch and a Hex 37 waveguide respectively (for a *F/7.6* beam). It is important to first point out that since these are far-field images, the axes are effectively plotting increasing *NA* as we move away from the origin. It is clear to see that although the probe light was injected with a near flat-top distribution of *NAs* (see cross-sectional line profiles in Fig. 6(a)) it leaves the waveguide in a near Gaussian distribution (see line profiles in Fig. 6(b)) which suggests both cross-coupling of the modes within the waveguide, as well as greater losses for the higher order modes (high *NAs*). Figures 6(c) and 6(d) reveal that for low injected *F/#s* the output *F/#s* are insensitive to changes in the injection conditions. This is because the light is being injected at lower *F/#* than the waveguide can support (i.e. above the cutoff *NA*) and hence changes to the injection in this region are not reflected at the output. The point where these data points intersect with the *output equals input line* yields the cutoff of the guides. The cutoff *F/#* and corresponding *NA* is summarized in table 1 below for 3 Hex 37 waveguides written with different translation speeds. The cutoff *F/#'s* are consistent, to within experimental uncertainty, to the point at which the throughputs flattened off in Fig. 5(a) and reach ~100% in Fig. 5(b), which means that the lower throughputs at smaller *F/#s* are due to light being injected above the cutoff of the guides. As the *NA* is related to the index contrast between the core and cladding by $NA = \sqrt{(n_{core}^2 - n_{clad.}^2)}$ for a step index waveguide, then the monotonically decreasing cutoff *NA* as a function of increasing translation speed for the set of Hex 37 guides indirectly validates our previous assertion; slower translation speeds increase the strength of the index modification (highlighted by the blue arrow in Fig. 5(b)). The *NAs* of the waveguides are relatively low when compared with standard MM fibers that have *NAs* in the range of 0.12 - 0.22. Indeed, if for simplicity we assume a step index profile then the waveguides have an effective refractive index contrast of 1.2-1.6 × $10^{-3}$, which is between one quarter and one third that for a standard SM fiber such as SMF-28e [21]. Nonetheless as we will show, this does not affect the use of such guides in astronomy.

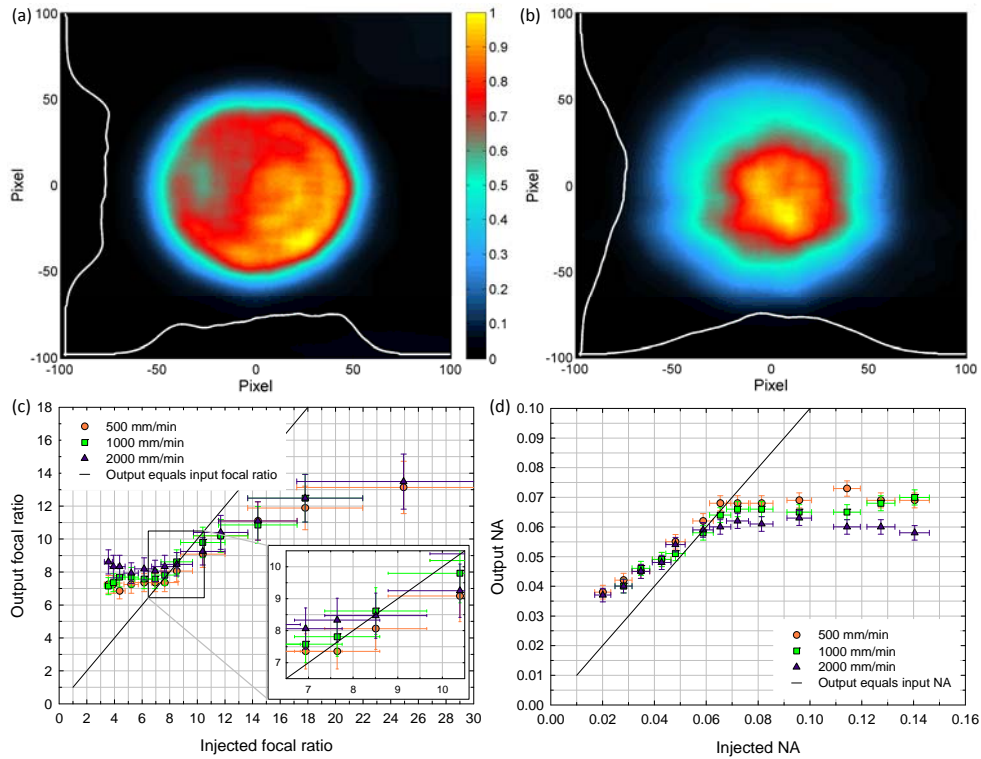

Fig. 6. Far-field energy distribution for the (a) launch and (b) Hex 37 waveguide inscribed with a translation speed of 1000 mm/minute for an injection of *F/7.6*. (c) Output focal ratio as a function of injected focal ratio and (d) Output *NA* as a function of the injected *NA* for Hex 37 waveguide structures written with 500, 1000 and 2000 mm/minute translation speeds. A line has been added which represents the *output equals input line* in order to highlight regions where performance degradation occurs. Inset in Fig. (c) is a magnified view of the center of the plot.

Table 1. Cutoff *F/#/NA* for Hex 37 waveguides written at various translation speeds.

| Waveguide type | Hex 37 | |
|---|---|---|
| Translation speed (mm/minute) | *F/#* | *NA* |
| 500 | 7.4±0.5 | 0.068±0.009 |
| 1000 | 8.0±0.5 | 0.062±0.008 |
| 2000 | 8.5±0.5 | 0.059±0.007 |

There are several other points of interest from the graphs in Fig. 6. Firstly, in the region between *F/*7.5 and F/10 the output *F/#* equals the injected *F/#* to within experimental uncertainty. Secondly, beyond ~*F/*10 it can be seen that the output *F/#* is lower than the injected *F/#* which means that the light diverges into a larger cone angle at the output than that corresponding to the injected light at the input. This phenomenon is called focal ratio degradation (FRD) and results from the various attributes of the waveguides (such as index profile inhomogeneities, stress at the facets due to polishing, etc) causing light to couple to higher order modes [22]. This unwanted property is present in MM optical fibers as well [22], however unlike MM fibers these waveguides offer a narrow window (*F/*7.5 to *F/*10) within which no degradation is observed. This is the first time, to the best of our knowledge, that FRD has been explored in optical waveguides. It is important to understand this effect, as it will affect the use of such guides both in isolation and in combination with photonic lanterns.

*3.2 Non-overlapped multimode waveguide structures - optimizing the pitch*

A set of Circ 19 and Hex 19 waveguide structures were inscribed at a fixed translation speed of 1000 mm/minute with pitches ranging from 8-15 μm. The normalized throughputs of the Circ and Hex 19 waveguide structures as a function of the injected *F/#* and the lattice pitch are summarized in Fig. 7 below. It should be noted that as the physical size of the waveguides varied throughout these sets, the windowing function used to compensate for the coupling loss (dashed grey line in Fig. 3) was scaled in proportion to the size of the guide. By casual inspection it is clear that the data sets for the two lattice geometries are similar in nature. In particular, the throughputs are approximately constant (to within experimental uncertainty) for lattice pitches between 8-10 μm and decrease monotonically for increasing lattice pitches. In addition the throughputs increase for increasing *F/#* as was the case for the Hex 37 structures, which is again a result of the fact that at low *F/#s* light is being injected above the cutoff of the waveguides. Interestingly for *F/#>*8, the throughput reaches ~100% (to within the 7% experimental uncertainty) for lattice pitches up to 10 μm. Of this narrow range of pitches only the 10 μm pitch is truly non-overlapped. This is a favorable result because it means that non-overlapped structures can be used without compromising on the throughput. It is clear that the throughputs are slightly lower for the Hex 19 lattices which is partially attributed to the fact that a circular shaped windowing function was used to rescale the launch field for all waveguides, which over estimates the light coupled into the Hex structures.

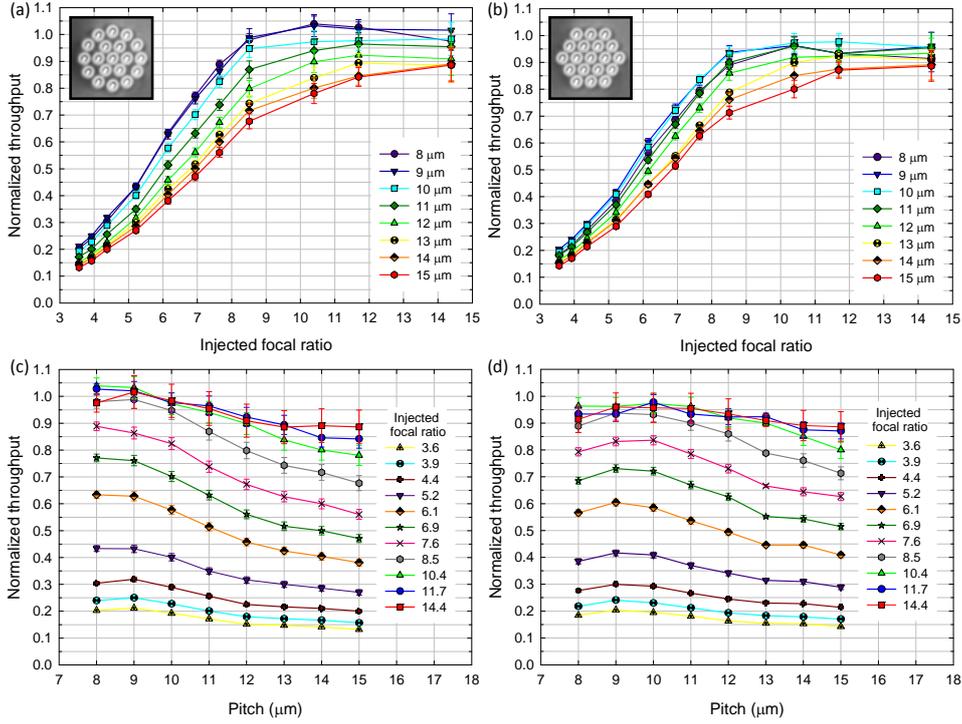

Fig. 7. Normalized throughput for various lattice pitches as a function of injected focal ratio for (a) Circ 19 and (b) Hex 19 waveguide structures. Normalized throughput for various injected focal ratios as a function of lattice pitch for (c) Circ 19 and (d) Hex 19 waveguide structures. The translation speed for inscription was fixed at 1000 mm/minute for all waveguides.

There are several other features of these figures that are best understood by reviewing some basic waveguide equations [23]. The approximate number of modes ($Nm$) supported by a step index MM waveguide, in the regime where $V \gg 2.405$ (the single-mode cutoff) is given by the following equation

$$Nm \approx \frac{V^2}{2} = \frac{1}{2}\left(\frac{\pi d}{\lambda} NA\right)^2 \qquad (2)$$

where $V$ is the normalized frequency, $d$ is the diameter of the waveguide and $\lambda$ is the wavelength of light. As the pitch is increased, and hence the diameter of the waveguide increases, Eq. 2 reveals that the number of modes supported by the guide will increase quadratically. The maximum number of modes supported has an upper limit that is reached when the SM tracks are separated sufficiently apart from one another so as to be completely de-coupled. In this case there will be $N$ modes, which correspond to the fundamental mode for each SM track ($N=19$ in our case, ignoring polarizations). Although there are more modes for a lattice with a greater pitch, the range of effective indices of the modes reduces to the point where they become completely degenerate when the SM tracks are sufficiently de-coupled (all contain identical fundamental modes). This means that the modes that are supported have a decreasingly narrower range of $k$ vectors and hence rays with large angles of inclination to the optic axis (high *NAs*) can no longer be supported; hence the *NA* of the guides decreases. This can be understood in an intuitive fashion by considering what happens as the SM tracks are pulled further and further apart; they have an ever-decreasing fill factor and hence the effective index contrast of the guide reduces, reducing the *NA*. It is the reduction in *NA* that accounts for the drop in throughput for increasing lattice pitches as seen in Fig. 7(c) and (d). This also explains why there is a smaller drop in throughput for slower beams (>*F/10*) as the pitch is increased.

*3.3 Non-overlapped multimode waveguide structures - throughput and cutoff focal ratio*

After determining that a lattice pitch of 10 μm results in non-overlapped structures with minimal affect on the throughput, the index contrast of the guides was varied in order to optimize the throughput at low *F/#* (i.e. maximize the cutoff *NA*), by inscribing a set of Circ and Hex 19 structures with a range of translation speeds. Figure 8 summarizes the throughput results. Once again there is a strong similarity between the results for the two lattice geometries, which is not unexpected and means that there is little performance benefit to choosing one over the other. Further, the maximum throughput of both waveguide types was ~100% at high *F/#*, a value consistent with that of the overlapped Hex 37 structure. The final point of interest is the steeper slope in the throughput curves in Figs. 8(c) and 8(d) for *F/#*<8 as compared to those for the overlapped Hex 37 structure presented in Fig. 5(b). This is because the increase in the size and index contrast of the SM tracks with decreasing translation speed has a more significant impact on a sparse lattice that consists of unmodified glass regions as opposed to a fully overlapped structure.

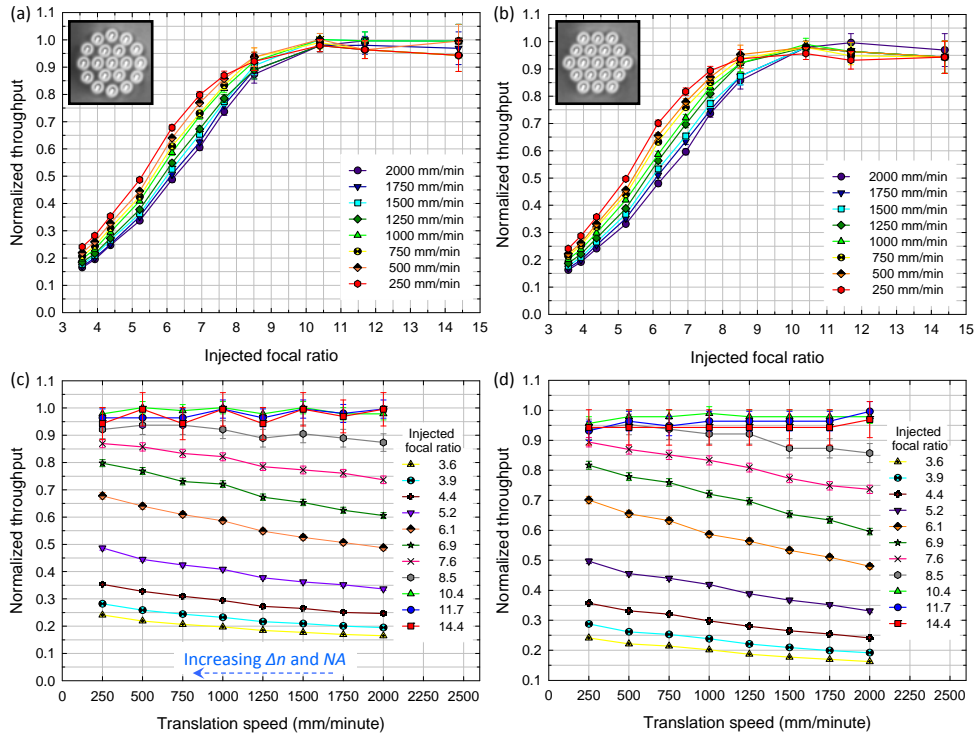

Fig. 8. Normalized throughput for various translation speeds of inscription as a function of injected focal ratio for (a) Circ 19 and (b) Hex 19 waveguide structures. Normalized throughput for various injected focal ratios as a function of translation speed used for inscription of the (c) Circ 19 and (d) Hex 19 waveguide structures.

The optical characteristics of three Circ 19 and Hex 19 waveguide structures are summarized in Fig. 9. From the figures the cutoff *F/#s* and *NAs* were determined and are summarized in Table 2 below. It can be seen that the cutoff *NA* for a given translation speed is consistent to within uncertainty for the two lattice geometries. Comparing the cutoff *NAs* for the Circ 19 and Hex 19 structures with that of the Hex 37 structure in section 3.1 above it is clear that the former are 15-20% lower. This is a result of the lower effective index of the guides due to the sparse nature of the lattices relative to the overlapped Hex 37 structure. Finally the region over which the waveguides do not exhibit FRD has also shifted from that of the Hex 37 structures and is now from *F/9* to *F/12* (to within the limits of experimental uncertainty) for both structure types, above which degradation is clearly evident.

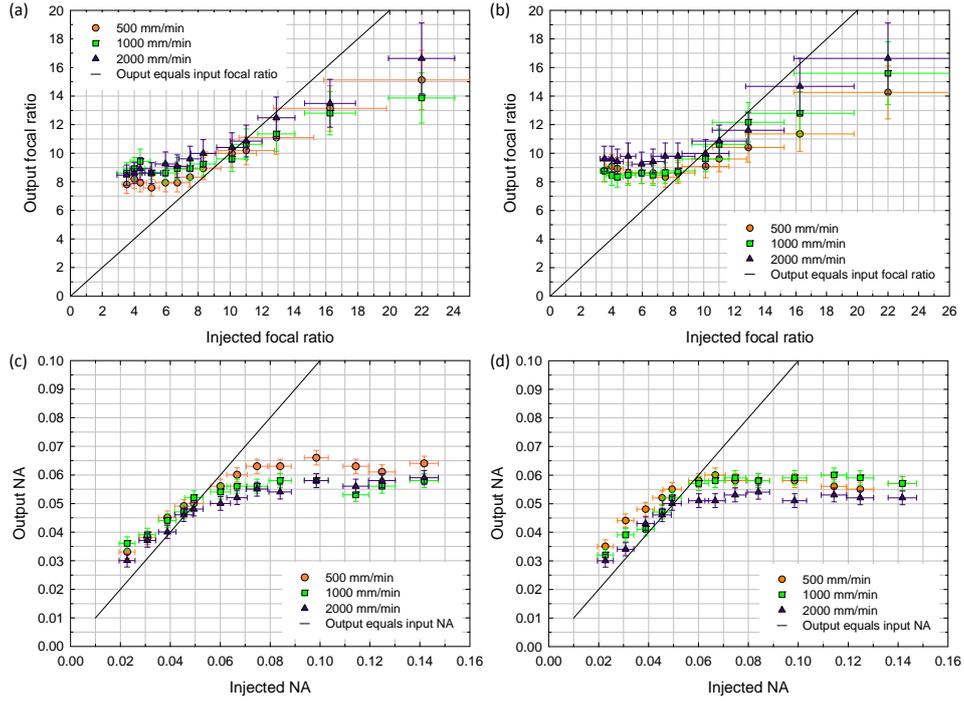

Fig. 9. Output focal ratio as a function of injected focal ratio for the (a) Circ 19 guides and the (b) Hex 19 guides. Output *NA* as a function of the injected *NA* for the (c) Circ 19 and the (d) Hex 19 waveguide structures written with 500, 1000 and 2000 mm/minute translation speeds. A line has been added which represents the *output equals input line* in order to highlight regions where performance degradation occurs.

Table 2. Cutoff *F/#/NA* for Circ and Hex 19 waveguides written at various translation speeds.

| Waveguide type | Circ 19 | | Hex 19 | |
|---|---|---|---|---|
| **Translation speed (mm/minute)** | *F/#* | *NA* | *F/#* | *NA* |
| 500 | 9.4±0.5 | 0.053±0.006 | 8.6±0.5 | 0.058±0.007 |
| 1000 | 9.5±0.5 | 0.053±0.006 | 8.8±0.5 | 0.057±0.006 |
| 2000 | 10±0.5 | 0.050±0.005 | 9.8±0.5 | 0.051±0.005 |

For completeness we show some near-field intensity distributions for the Circ and Hex 19 structures for two different translation speeds of inscription in Fig. 10 below. It can be seen that in all cases the energy is localized around the regions with the highest index contrast (i.e.

the position of the SM tracks). From the images it is clear that all SM tracks are coupled, as there is a strong and continuous field between all SM tracks. Finally the intensity is more uniform in the case of the lattices written at high translation speeds ((b) and (d)) which is the result of the lower index contrast of the SM tracks as explained previously which allows more cross-coupling between the constituent tracks.

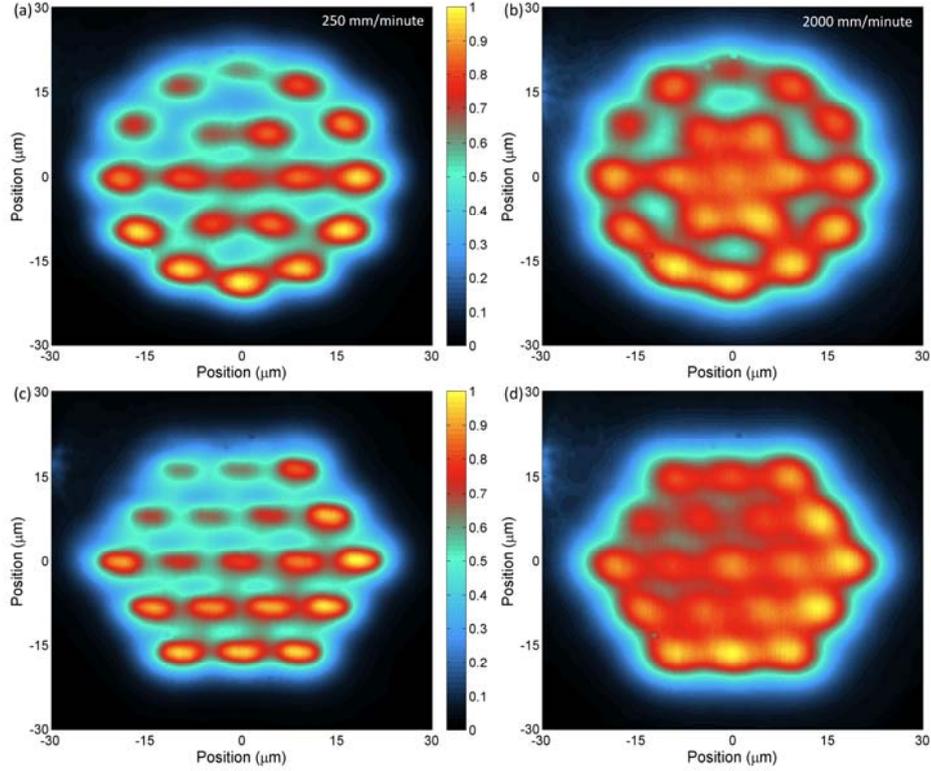

Fig. 10. Near-field intensity distributions for Circ 19 structures ((a) and (b)) and Hex 19 structures ((c) and (d)), for two translation speeds of inscription; 250 mm/minute ((a) and (c)) and 2000 mm/minute ((b) and (d)).

## 4. Composite multimode waveguides for slit reformating devices for diffraction-limited spectrographs

The previous section has shown that composite MM waveguides constructed from overlapping and non-overlapping SM tracks can transmit all of the injected light over a 30 mm length of chip if injected below the cutoff of the guides. Further, we have identified regions where the waveguide structures exhibit FRD and a narrow window where they don't exhibit any FRD at all. These optical characteristics are favorable and this section will outline how such waveguide structures could be utilized in next generation astronomical instrumentation.

As the cutoff *NA* of the waveguides is low (maximum of 0.068 from Table 1) compared to even standard SM fibers (0.13), bend losses in such guides would be substantial. This limits the use of the current MM guides to devices that consist of straight sections only, which rules out a number of applications. It should be noted that it is possible to increase the index contrast by at least 10-20% by further optimizing the writing pulse energy, which would help alleviate some of the bend loss issue if it is required.

The leading application for composite MM waveguide structures in astronomical instrumentation is as a slit reformating device for diffraction-limited spectrographs. In principle such a device would first collect the seeing-limited light directly from the focal plane of a telescope into a composite MM waveguide structure. It would then undergo a lantern-like transition to redistribute the field amongst many SM waveguides before finally being reformatted into a linear array of closely spaced guides that constitute the slit of a spectrograph; as conceptually illustrated in Fig. 11. For clarity, although the proposed device utilizes bends, they are at a point along the device where the light is guided by the isolated SM tracks which have *NA*~0.12, commensurate with standard SM fibers (0.13). The greater NAs of the SM tracks in isolation means that they suffer from significantly less bend losses than the composite MM structures do, and hence the bends required by the structure will not be an issue. More rigorous calculations will be offered in a future publication. There are several considerations for designing and implementing such a device.

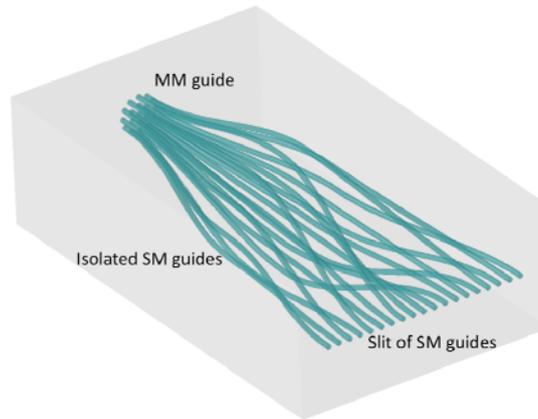

Fig. 11. Concept for a slit reformatting device for a diffraction-limited spectrograph. Three key regions are highlighted; the MM guide that collects the seeing-limited light from the focus of the telescope, the isolated SM guides and the slit formed from the SM guides.

*4.1 Type of waveguide to be used*

In order to minimize scattering losses within the lantern-like transition in the proposed device, it is important to have a homogenous refractive index contrast over all elements within the transition. However, this is difficult to achieve in practice when using the overlapped structure geometry as multiple overpasses of a region with the writing laser leads to a locally higher index change that is difficult to homogenize. In this regard the non-overlapping geometries demonstrated above, namely the Circ and Hex 19 structures are the ideal choice because no region of the glass substrate is processed by the inscribing laser more than once.

*4.2 Injecting light from the telescope*

For optimum device performance it is important to carefully consider how the light is injected from the telescope into the device. The first step is to inject the light as close to cutoff as possible (*F/8*-*F/9*). This will ensure that the physical size of the seeing-limited point-spread function (PSF) will be minimized which means that the composite MM waveguide will require the minimum number of SM tracks possible. This can be achieved by choosing a focal plane with the appropriate *F/#* (i.e. primary ~*F/3*, Cassegrain ~*F/8*, Nasmyth ~*F/18*, etc) or alternatively by placing a micro-lens in front of the device to alter the speed of a given focus. If we take the Anglo-Australian Telescope (AAT) for example, an astronomical facility renowned for commissioning next-generation technology such as the device proposed here, the Cassegrain focus will deliver a *F/8* beam which best matches the cutoff of the waveguides presented. Since the median seeing at the site of the telescope is 1.6 arcseconds, the FWHM of the PSF in this focal plane is 186 μm. If a device based on a hexagonal lattice was placed in this plane the number of waveguides would need to be increased to 130 or 200 with lattice pitches of 15 or 10 μm respectively, in order to optimally sample the PSF. By using a larger pitch (15 μm) it is possible to minimize the number of waveguides required for the slit (130), which in turn minimizes the length of the slit and the overall detector size requirements. However, as can be seen in Fig. 7(b) and (d), there is a throughput penalty associated with using larger pitches and a hence a compromise must be found.

On the other hand for smaller observatories with ~ 0.4 m telescopes such as the one at Macquarie University Observatory, the 2.5 arcsecond seeing offers a 50 μm size PSF with a *F/10* beam. Such optical characteristics are ideal for injecting into the proposed slit-reformatting device based on the composite MM waveguides fabricated and characterized in this study.

*4.3 Fiber feeds and fore-optics*

It would also be possible to place the slit reformating device and spectrograph in a gravitationally invariant environment (stable, none tilting environment) adjacent to the telescope itself and transport the light to the reformating device via a MM optical fiber. In designing such a system it is imperative that the FRD of the transport fiber be considered. Indeed it is not possible to transport light in standard MM fibers while maintaining high *F/#s*. This means that the low *F/#* output of the MM fiber would need to be reimaged with fore-optics in order to slow it down before it could be injected efficiently into the low *NA* composite waveguide structures. In essence choosing this method will always result in a larger number of SM tracks required for the device.

The final consideration in designing an optimized slit-reformating device is the number of modes supported by each section. This point has several considerations that will be discussed in detail in a follow-up publication.

**Conclusion**

In this body of work we have demonstrated the realization of composite MM waveguide structures, which were fabricated by ultrafast laser inscription. The MM waveguides consisted of a lattice of SM tracks, which were both overlapped and non-overlapped in nature. Upon testing at 1550 nm, it was determined that both the overlapped and non-overlapped structures have negligible propagation losses for injected focal ratios >8. As expected the overlapped structures had higher cutoff *NAs* ($0.065 \pm 0.005$) than the non-overlapped structures ($0.055 \pm 0.005$), a result attributed to the higher effective index contrast of the waveguides. It was determined that near, but below the cutoff of the waveguides there was a narrow region where light propagating through the guides would retain the focal ratio at the output with which it was injected. For larger injected focal ratios focal ratio degradation was observed.

Although the low *NAs* prevent such guides from being used in isolation in their current form, our analysis indicates that such structures can form the building blocks for slit reformating devices for diffraction-limited spectrographs. Reformating of this nature would empower smaller telescopes (<0.5 m) to be able to conduct simple, high-resolution spectroscopic surveys efficiently. This would alleviate to some extent the demand on the larger world-class astronomical observatories for such mundane tasks, reserving their precious time for the detailed follow up studies.

**Acknowledgements**

The authors would like to acknowledge informative discussions with Dr Anthony Horton. This research was conducted with the Australian Research Council Centre of Excellence for Ultrahigh Bandwidth Devices for Optical Systems (project number CE110001018) and the assistance of the LIEF and Discovery Project programs. This work was supported by the OptoFab node of the Australian National Fabrication Facility (ANFF).